\documentclass[aps,prl,reprint,showpacs,superscriptaddress]{revtex4-1}
\usepackage{graphicx}
\usepackage{miller}
\usepackage{wasysym}
\usepackage{xspace}
\usepackage{braket}
\usepackage{siunitx}
\usepackage{hyperref} 
\usepackage{float}
\usepackage{textcomp}

\usepackage{silence}
\DeclareSIUnit\electrons{e\textsuperscript{--}}
\DeclareSIUnit\neutrons{neutrons}
\DeclareSIUnit\ppm{ppm}
\DeclareSIUnit\ppb{ppb}
\DeclareSIUnit\lines{l}
\DeclareSIUnit{\calorie}{cal}

\newcommand{\oneelec}[1]{\ensuremath{\mathrm{#1}}}

\newcommand{\spinstate}[3]{\ensuremath{{}^{#1}\mathrm{#2_{#3}}}}

\newcommand{\Cthir}{\ensuremath{^{13}\mathrm{C}}\xspace}

\newcommand{\NVnb}{\ensuremath{\mathrm{NV}}\xspace}

\newcommand{\NVminus}{\ensuremath{\NVnb^{-}}\xspace}

\newcommand{\NV}{\NVnb\xspace}

\newcommand{\SiVneutral}{\ensuremath{\mathrm{SiV}^{0}}\xspace}
\newcommand{\SiVminus}{\ensuremath{\mathrm{SiV}^{-}}\xspace}
\newcommand{\SiV}{\ensuremath{\mathrm{SiV}}\xspace}

\newcommand{\Trigonal}{\ensuremath{\mathrm{C}_{\mathrm{3v}}}\xspace}
\newcommand{\DIIId}{\ensuremath{\mathrm{D_{3d}}}\xspace}

\setcitestyle{numbers,square}

\newenvironment{packed_itemize}
{
\begin{itemize}
\setlength{\itemindent}{-2pt}
\setlength{\itemsep}{+2pt}}
{\end{itemize}}

\begin{document}
\title{The neutral silicon-vacancy center in diamond: spin polarization and lifetimes}

\author{B.\ L.\ Green}
\email{b.green@warwick.ac.uk}
\altaffiliation{Corresponding Author}
\author{S.\ Mottishaw}
\author{B.\ G.\ Breeze}
\affiliation{Department of Physics, University of Warwick, Coventry, CV4 7AL, United Kingdom}
\author{A.\ M.\ Edmonds}
\affiliation{Element Six Limited, Global Innovation Centre, Fermi Avenue, OX11 0QR, United Kingdom}
\author{U.\ F.\ S.\ D'Haenens-Johansson}
\affiliation{Gemological Institute of America, 50 W 47\textsuperscript{th} St, New York, NY 10036, United States of America}
\author{M.\ W.\ Doherty}
\affiliation{Laser Physics Centre, Research School of Physics and Engineering, Australian National University, Australian Capital Territory 0200, Australia}
\author{S.\ D.\ Williams}
\author{D.\ J.\ Twitchen}
\affiliation{Element Six Limited, Global Innovation Centre, Fermi Avenue, OX11 0QR, United Kingdom}
\author{M.\ E.\ Newton}
\email{m.e.newton@warwick.ac.uk}
\affiliation{Department of Physics, University of Warwick, Coventry, CV4 7AL, United Kingdom}

\begin{abstract}
We demonstrate optical spin polarization of the neutrally-charged silicon-vacancy defect in diamond (\SiVneutral{}), an $S=1$ defect which emits with a zero-phonon line at \SI{946}{\nano\meter}. The spin polarization is found to be most efficient under resonant excitation, but non-zero at below-resonant energies. We measure an ensemble spin coherence time $T_2\SI{>100}{\micro\second}$ at low-temperature, and a spin relaxation limit of $T_1>\SI{25}{\second}$. Optical spin state initialization around \SI{946}{\nano\meter} allows independent initialization of \SiVneutral{} and \NVminus{} within the same optically-addressed volume, and \SiVneutral{} emits within the telecoms downconversion band to \SI{1550}{\nano\meter}: when combined with its high Debye-Waller factor, our initial results suggest that \SiVneutral{} is a promising candidate for a long-range quantum communication technology.
\end{abstract}

\maketitle

Point defects in diamond have attracted considerable interest owing to their application for quantum information processing, communication, and metrology. The most-studied defect, the negatively-charged nitrogen-vacancy (\NVminus) center, possesses efficient optical spin polarization and spin-state dependent fluorescence, enabling its exploitation as an ultra-sensitive nano-scale magnetic field sensor \cite{Doherty2013,Rondin2014,Maletinsky2012}. However, the zero phonon line (ZPL) of \NVminus{} accounts for only a few percent of its total emission \cite{Faraon2012}, leading to low efficiency in coherent photonic applications. The negatively-charged silicon-vacancy (\SiVminus) center has also received significant interest as its high Debye-Waller factor ($\approx0.8$ \cite{Neu2011}) makes it an attractive candidate for long-range quantum computation and communication. However, the exceptional optical properties of \SiVminus are not matched by its spin properties, where a large spin-orbit coupling in the ground state enables phonon-assisted spin-state depopulation, resulting in spin-lattice relaxation-limited coherence lifetimes of \SI{40}{\nano\second} even at \SI{5}{\kelvin} \cite{Becker2016}: efforts are ongoing to overcome this limitation by strain engineering, but currently liquid helium temperatures and below are required to access and readout \SiVminus spin states \cite{Pingault2017}. 

\begin{figure}[tb]
	\centering
	\includegraphics[width=\columnwidth]{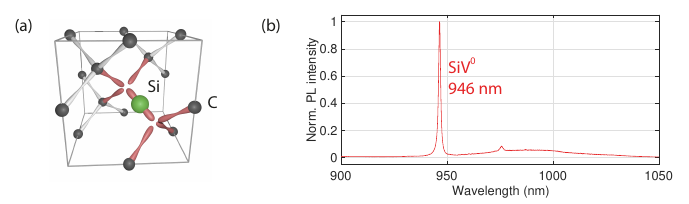}
	\caption{(a) The structure of the silicon-vacancy defect in the diamond lattice: the silicon adopts a bond-center location, leading to \DIIId{} symmetry (\hkl<111> axis) in both the neutral and negative charge states \cite{Goss1996}. (b) Photoluminescence spectrum of \SiVneutral{} (zero phonon line at \SI{946}{\nano\meter}) at a temperature of \SI{80}{\kelvin} in sample~B. The feature at \SI{976}{\nano\meter} is related to \SiVneutral{} and may indicate local strain \cite{Gali2013}.}
	\label{fig:structure_and_optical_absorption}
\end{figure}

The neutrally-charged silicon-vacancy (\SiVneutral{}) has a ground state electron spin $S=1$. Unlike the \NV{} center, where the nitrogen remains covalently bonded to three carbon atoms and the nitrogen-vacancy axis forms a \Trigonal{} symmetry axis, the silicon atom in \SiV{} adopts a bond-center location, with a \DIIId{} axis formed by the \hkl<111> joining the split-vacancy [Fig.~\ref{fig:structure_and_optical_absorption}(a)]. \SiVneutral{} has been characterized both by electron paramagnetic resonance (EPR) \cite{Iakoubovskii2001a,Edmonds2008a} and optical absorption/photoluminescence (PL) \cite{DHaenens-Johansson2011}. Similarly to \SiVminus{}, the neutral charge state also has a high Debye-Waller factor, with the majority of its photons emitted at the primary zero-phonon line (ZPL) at \SI{946}{\nano\meter} (\SI{1.31}{\electronvolt}) [Fig.~\ref{fig:structure_and_optical_absorption}(b)]: this transition has been shown to occur between a \spinstate{3}{A}{2g} ground state (GS) and \spinstate{3}{A}{1u} excited state (ES) \cite{DHaenens-Johansson2011,Gali2013}. Quenching of PL at low temperature indicates the presence of a shelving state \SI{5}{\milli\electronvolt} below the ES \cite{DHaenens-Johansson2011}. The zero-field splitting (ZFS) in the GS is $D=\SI[retain-explicit-plus]{+1000}{\mega\hertz}$ at \SI{300}{\kelvin} \cite{Edmonds2008a}. The ZFS is highly temperature-dependent, being approximately linear in the range \SIrange{50}{150}{\kelvin} with $dD/dT=\SI{-337}{\kilo\hertz\per\kelvin}$, and an average of $\SI{-202}{\kilo\hertz\per\kelvin}$ between \num{50} and \SI{300}{\kelvin} --- these values are significantly higher than for \NVminus{} at $\SI{-74}{\kilo\hertz\per\kelvin}$ \cite{Acosta2010b}. Finally, non-equilibrium populations of the four symmetry-related defect orientations have been observed in grown-in \SiVneutral{} when the diamond crystal is grown on substrates of particular crystallographic orientation \cite{DHaenens-Johansson2011}. This behaviour has been previously observed in \NVminus{} \cite{Edmonds2012,Lesik2014} and \SiVminus{} \cite{Rogers2014a}.

\begin{figure}[tbp]
	\centering
	\includegraphics[width=0.95\columnwidth]{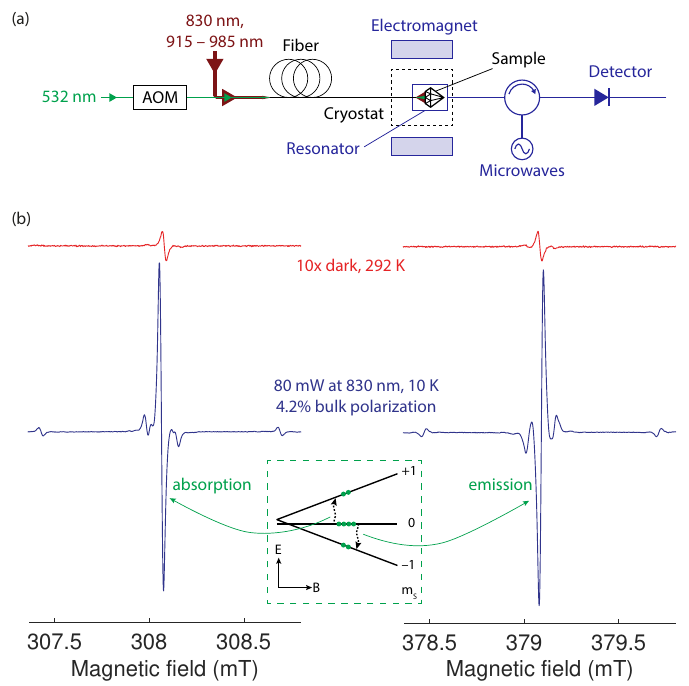}
	\caption{(a) Schematic of optically-pumped electron paramagnetic resonance (EPR) measurements. Light from a laser is delivered via optical fiber to the sample mounted into the cryostat. CW measurements performed using lasers at \num{532}, \num{830}, \SI{915}{\nano\meter} and a tuneable \SIrange{915}{985}{\nano\meter}. The \SI{532}{\nano\meter} laser is switched using an acousto-optic modulator (AOM) for pulsed EPR measurements. Components of the EPR spectrometer are in blue. (b) EPR spectra of the $m_{s}=0\leftrightarrow+1$ (left) and $m_{s}=0\leftrightarrow-1$ (right) transitions of \SiVneutral{} in sample~A, collected with applied magnetic field $B$ within \SI{2}{\degree} of \hkl<111>. Upper spectra collected at room temperature; lower spectra collected at a sample temperature of \SI{10}{\kelvin} during optical pumping with \SI{80}{\milli\watt} (\SI{10.2}{\watt\per\centi\meter\squared}) at \SI{830}{\nano\meter}. The opposite phase of the low- and high-field lines indicate enhanced absorption and emission, respectively: we achieve a bulk polarization of $\xi=\SI{4.2}{\percent}$. Optically-pumped spectra have been offset for clarity; magnetic field given for room temperature measurement. Inset: schematic of ground-state energy levels with nominal spin-polarized populations illustrating origin of enhanced absorption and emission.}
	\label{fig:epr_spectra}
\end{figure}

We have studied \SiVneutral{} in two samples grown by chemical vapor deposition (CVD); silicon was introduced by adding silane to the process gasses during growth. Sample~A was grown on a \hkl{100}-oriented high pressure high temperature (HPHT) substrate, irradiated to a dose of \SI{5.4E17}{\electrons\per\centi\meter\squared} at \SI{2.0}{\mega\electronvolt} and annealed for four hours each at \num{400} and \SI{800}{\celsius} to produce \SI{5+-2}{\ppb} of \SiVneutral{}. Sample~B was grown on a \hkl{113}-oriented HPHT substrate. The sample was irradiated with \SI{1.5}{\mega\electronvolt} electrons to a dose of \SI{1E18}{\electrons\per\centi\meter\squared} before annealing for \SI{4}{\hour} at \SI{900}{\celsius} to produce \SI{75+-8}{\ppb} of \SiVneutral{}. 

To investigate the behavior of \SiVneutral{} under optical excitation, we perform both CW and pulsed EPR measurements (Bruker~E580 spectrometer) using a dielectric resonator (Bruker ER~4118X-MD5) and cryostat (Oxford Instruments CF935) for variable temperature measurements. Optical excitation from various laser sources is delivered to the sample via $\diameter\SI{1}{\milli\meter}$ core fiber held in place with a rexolite rod [Fig.~\ref{fig:epr_spectra}(a)]. For pulsed measurements the \SI{532}{\nano\meter} laser (CNI MGLIII532) is switched using an acousto-optic modulator. Quantitative EPR measurements are carried out using non-saturating microwave powers.


Figure~\ref{fig:epr_spectra}(b) illustrates the effect of in-situ continuous optical pumping at \SI{830}{\nano\meter} on the EPR spectrum of \SiVneutral{}. The low- and high-field resonances spin-polarize into enhanced absorption and emission, respectively, under optical pumping. The ZFS of \SiVneutral{} is known to be positive \cite{Edmonds2008a} (i.e. $\ket{m_{s}}=\pm 1$ are higher in energy than $\ket{m_{s}}=0$ at zero magnetic field) and thus the low- and high-field resonances correspond to the $m_{s}=0\leftrightarrow+1$ and $m_{s}=0\leftrightarrow-1$ transitions. Therefore, the optical pumping is generating enhanced population in the $m_{s}=0$ state [Fig.~\ref{fig:epr_spectra}, inset] --- analagous to the polarization behaviour observed in \NVminus{} when excited with light of wavelength $\leq\SI{637}{\nano\meter}$. Qualitatively similar polarization is observed in both samples for excitation at \num{532} (\SI{2.33}{\electronvolt}), \num{830} (\SI{1.49}{\electronvolt}) and \SI{915}{\nano\meter} (\SI{1.36}{\electronvolt}) at both \num{80} and \SI{10}{\kelvin} \cite{Note1}. Given the high Debye-Waller factor of \SiVneutral{} [Fig.~\ref{fig:structure_and_optical_absorption}(b)], it is surprising that spin polarization is generated over such a wide energy range. Photoconductivity measurements of diamond containing \SiVneutral{} indicate a strong photocurrent at \SI{830}{\nano\meter} \cite{Allers1995} and hence charge effects are expected to be important for excitation at \SI{830}{\nano\meter} and below: the polarization may therefore be a result of the capture of a hole (electron) at $\mathrm{SiV^{-}}$ ($\mathrm{SiV^{+}}$), and not intrinsic to \SiVneutral{}. Theoretical studies indicate that green excitation may also excite from deep valence-band states \cite{Gali2013}.

\begin{figure}[bth]
	\centering
	\includegraphics[width=0.95\columnwidth]{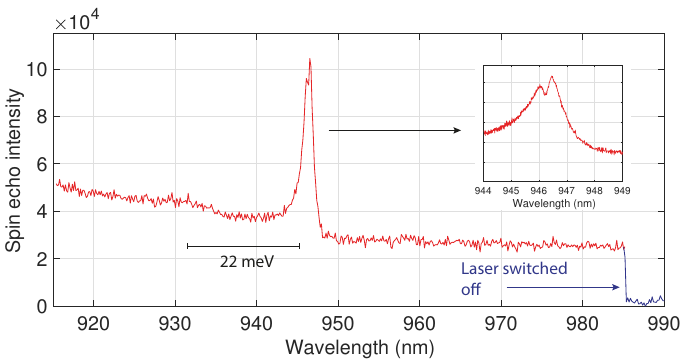}
	\caption{Spin polarization of sample~A measured by a two-pulse Hahn echo while sweeping the incident laser wavelength at an optical power of \SI{9}{\milli\watt}. The spin polarization decreases before reaching a sharp peak at the ZPL wavelength. Above-ZPL excitation continues to generate spin polarization, with a sharp signal decrease observed when the laser is switched off. Inset: high-resolution scans over the ZPL indicate that the ZPL is split by approximately \SI{0.4}{\nano\meter} ($\approx\SI{134}{\giga\hertz}$).} 
	\label{fig:ple}
\end{figure}

We investigate the possibility of an internal spin polarization mechanism by performing a Hahn echo-detected optical frequency-swept measurement: the optical frequency of a widely-tuneable narrow-linewidth laser (TOPTICA CTL 950) is swept over the ZPL, and we detect the resulting EPR enhancement by a two-pulse Hahn echo measurement at each frequency [Fig.~\ref{fig:ple}]. A sharp increase in polarization is observed for resonant excitation at the ZPL (\SI{946}{\nano\meter}), confirming a spin-polarization mechanism internal to \SiVneutral{} and unambiguously identifying the \SI{946}{\nano\meter} ZPL with \SiVneutral{}. In stark contrast to the behavior of \NVminus{}, polarization is observed even at sub-ZPL energies. Additionally, high-resolution measurements reveal a small ZPL splitting of \SI{0.4}{\nano\meter} ($\approx\SI{134}{\giga\hertz}$) [Fig.~\ref{fig:ple}, inset]. The origin of this splitting is unclear: optical measurements place the ZPL as a transition between states with no orbital degeneracy ($\spinstate{3}{A}{2g}\leftrightarrow \spinstate{3}{A}{1u}$) \cite{DHaenens-Johansson2011}, and orientational degeneracy is removed by measuring only those defects with their \hkl<111> axis parallel to the magnetic field. The effect of strain on a pair of orbital singlets is simply to shift the transition energy \cite{Mohammed1982}, and hence the observed splitting may indicate two populations of defects in distinct strain environments. Alternatively, if the excited state is in fact \spinstate{3}{E}{u}, then Jahn-Teller \cite{Hepp2014a} and spin-orbit effects become significant (excited state spin-orbit splitting in \SiVminus{} is approximately \SI{250}{\giga\hertz} \cite{Muller2014}): further investigation is required to determine the microscopic origin of the observed splitting. 

We define the degree of spin polarization as $\xi=\SI{100}{\percent}$ when all spins are in the $m_s=0$ state; and $\xi=\SI{0}{\percent}$ at thermal equilibrium (see \footnote{\label{footnote:suppInfo}See Supplemental Material at http://abc for details on spin polarization efficiency calculation, linewidth broadening at high temperature and raw polarized spectra for different excitation wavelengths.} for details of calculation). In both samples $\xi$ is found to increase from a typical $\SI{0.1}{\percent}$ at room temperature to approximately \SI{4}{\percent} at \num{80} and \SI{10}{\kelvin} for the same excitation \cite{Note1}. Maximum bulk polarization of $\SI{5.2}{\percent}$ is observed at \SI{10}{\kelvin} when pumping with \SI{80}{\milli\watt} at \SI{532}{\nano\meter}; maximum per-photon efficiency is found in sample~A under resonant ZPL excitation. At all temperatures and wavelengths the polarization is linear in optical power up to the maximum available at the sample, and therefore we neglect two-photon processes in our analysis.

\begin{figure}[t]
	\centering
	\includegraphics[width=0.95\columnwidth]{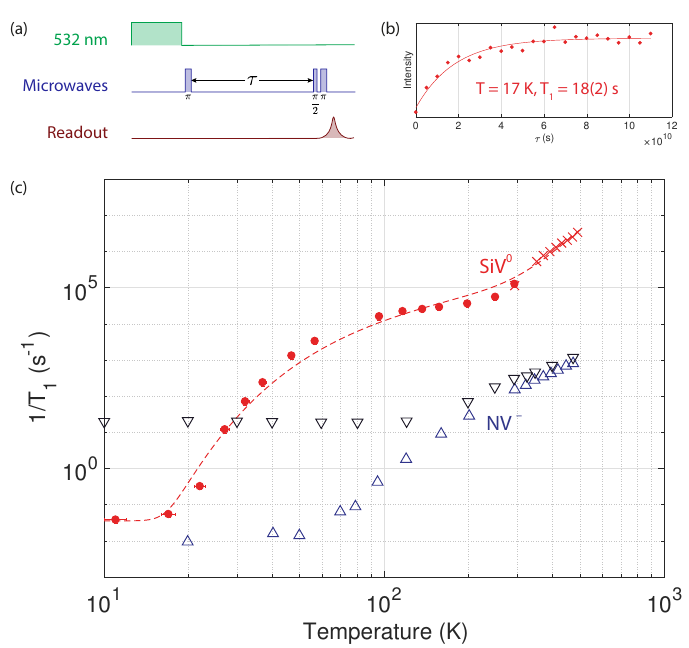}
	\caption{(a) Pulse sequence used to measure $T_1$. \SiVneutral{} is spin polarized by optical pumping at \SI{532}{\nano\meter} to increase the signal strength, then an echo-detected inversion-recovery measurement is performed to measure $T_1$. (b) Echo-detected inversion-recovery measurement at \SI{17}{\kelvin}, yielding a longitudinal spin lifetime of \SI{18+-2}{\second}. (c) Temperature dependence of $T_1$ in \SiVneutral{} (dots) and \NVminus{} (triangles, two different concentrations --- taken from \cite{Jarmola2012}). The \SiVneutral{} spin-lattice relaxation lifetime $T_1$ depends strongly on temperature in this sample. At low temperatures, $T_1\SI{>25}{\second}$. Crosses obtained via an indirect measurement of $T_1$ using linewidth (see \cite{Note1}); dashed line is a fit to different temperature-dependent relaxation processes (see text).}
	\label{fig:t1}
\end{figure}

The increase of $\xi$ with decreasing temperature may arise from several sources: temperature-dependent effects within the intrinsic spin-polarization mechanism itself can alter the polarization efficiency; and increases in electron longitudinal (spin-lattice) lifetime $T_1$ can lead to a greater macroscopic build-up of polarization for the same polarization efficiency. The latter was measured directly using the pulse sequence given in Figure~\ref{fig:t1}(a). The sample was placed in a magnetic field applied within \SI{2}{\degree} of \hkl<111> at a field strength of \SIrange{311}{316}{\milli\tesla} at \SIrange{292}{10}{\kelvin}: the changes in field are a result of the temperature-dependence of both the microwave resonator frequency and the ZFS. The sample was subjected to a polarizing \SI{1.5}{\milli\second} optical pulse to polarize into the $m_s=0$ state, and an inverting $\pi$ pulse of duration \SI{28}{\nano\second} was applied to the $m_{s}=0\leftrightarrow+1$ transition, transferring polarization into $m_{s}=+1$. After a variable delay $\tau$, the remaining spin polarization was measured using a Hahn echo detection sequence: this yields an exponentially-decaying signal with a single time constant equal to the longitudinal lifetime $T_1$ [Fig.~\ref{fig:t1}(b)]. 

The measured $T_1$ for \SiVneutral{} in sample~B are highly temperature-dependent [Fig.~\ref{fig:t1}(c)]: unlike \NVminus{}, which retains $T_{1}\sim\mathrm{ms}$ at room temperature, \SiVneutral{} lifetimes in this sample decrease from approximately \SI{25}{\second} at \SI{15}{\kelvin} to \SI{80}{\micro\second} at room temperature. EPR linewidth broadening is observed above room-temperature \cite{Note1}, and can be used as an indirect measure of $T_1$ in the limit that $T_2\leq 2T_1$ \cite{Slichter1990}, indicating $1/T_1\gtrsim\SI{1}{\mega\hertz}$. The dramatic temperature-dependence of $T_1$ is expected to account for the poor polarization efficiencies observed at room temperature. Interpretation of spin-lattice lifetimes in solids typically perfomed in terms of contributions from different phonon processes \cite{Abragam1986}. Interactions with single phonons (the so-called direct process, $1/T_{1}\propto T$) can be neglected, as the spin energies involved at X-band $T=\SI{10}{\giga\hertz}/k_B=\SI{0.5}{\kelvin}$ are at least an order of magnitude lower than the lowest measurement temperature (\SI{11}{\kelvin}). Relaxation via two phonons of different energies --- a Raman process --- can occur if the energy difference is equal to the spin transition energy, and takes the form $1/T_{1}\propto T^n$ where $n\in\{5,7,9\}$ depending on the spin levels involved, with $n=7$ typical for a non-Kramers doublet \cite{Abragam1986,Schweiger2001}. Finally, the Orbach process describes interaction with an excited spin state at a phonon-accessible energy $\Delta E$ above the ground state: the spin is excited by absorption of a phonon of energy $\hbar \Omega_{a} = \Delta E$ and relaxes to a different ground spin state by emission of a phonon $\hbar \Omega_{e} \neq \Delta E$. The $T_1$ data were therefore phenomenologically modeled using 
$$
\frac{1}{T_{1}} = A_{const}+A_{Raman}T^{7}+\frac{A_{Orbach}}{e^{{\Delta E}/k_BT}-1} \;.
$$
The fit in Figure~\ref{fig:t1}(c) was generated using the coefficients $A_{const}=\SI{0.036}{\per\second}$, $A_{Raman}=\SI{5.0E-13}{\second^{-1}\kelvin^{-7}}$, $A_{Orbach}=\SI{1.5E5}{\per\second}$ and $\Delta E=\SI{22}{\milli\electronvolt}$. The energy $\Delta E=\SI{22}{\milli\electronvolt}$ matches the phonon sideband observed in the echo-detected PLE measurement [Fig.~\ref{fig:ple}] and is close to the dominant phonon frequency $\hbar \Omega=\SI{28}{\milli\electronvolt}$ estimated from optical absorption measurements \cite{DHaenens-Johansson2011}: we therefore conclude that the primary phonon coupling frequency is similar in both the ground and excited states.  Multifrequency measurements would enable confirmation of the involved $T_1$ processes via their magnetic field dependence \cite{Abragam1986}. 

\begin{figure}[b]
	\centering
	\includegraphics[width=\columnwidth]{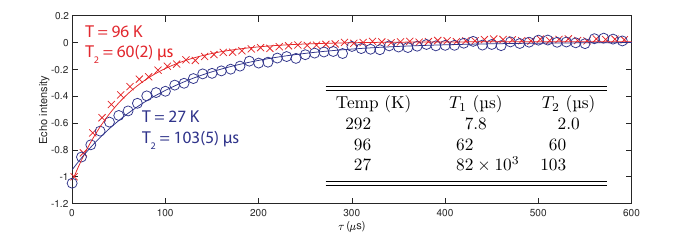}	
	\caption{$T_2$ decoherence lifetimes measured by echo decay at \num{96} and \SI{27}{\kelvin}. Inset: comparison of $T_1$ and $T_2$ spin-lifetimes. The measured $T_2$ is effectively limited by $T_2\leq2T_1$ at room temperature and \SI{96}{\kelvin}, but has reached a limit of \SI{98}{\micro\second} at \SI{27}{\kelvin}.}
	\label{fig:t2}
\end{figure}

The spin coherence time, $T_2$, is a critical parameter for many applications in sensing and quantum computation \cite{Dolde2013,Rondin2014,Zaiser2016}. We measure the $T_2$ directly using a Hahn echo-decay sequence ($\pi/2-\tau-\pi-\tau-echo$) \cite{Schweiger2001}, and find that $T_2$ changes from \SI{2.0}{\micro\second} at room temperature to \SI{103}{\micro\second} at \SI{27}{\kelvin} [Fig.~\ref{fig:t2}]. At both \num{292} and \SI{96}{\kelvin}, we find $T_2\approx T_1$, confirming that we are in the limit $T_2\leq2T_1$ \cite{Slichter1990}: at \SI{27}{\kelvin}, $T_2=\SI{103}{\micro\second}$ is limited by spin-spin interactions rather than $T_1=\SI{82}{\milli\second}$. This value is comparable to \NVminus{}, where ensemble measurements reach \SI{630}{\micro\second} without the use of decoupling sequences \cite{Stanwix2010}.

\begin{figure}[t]
	\centering
	\includegraphics[width=0.9\columnwidth]{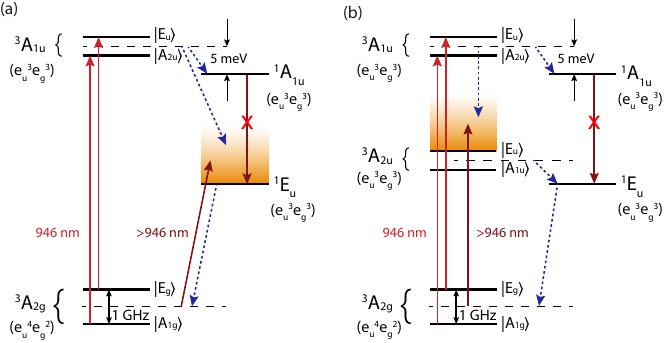}
	\caption{Possible mechanisms for generation of spin polarization into the GS $m_{s}=0$ state based on intersystem crossing (ISC) within \SiVneutral{}. Optical transitions are given as solid arrows; non-radiative transitions are dotted. The molecular orbital configuration for each state is given in brackets, spin-orbit (SO) states are given and spin-spin effects have been neglected. (a) Spin polarization occurs by spin-orbit coupling between the ES and a singlet state; below-ZPL polarization is generated by a dipole-allowed transition from the GS to the singlet which becomes weakly allowed due to SO mixing in the GS. (b) Spin polarization is generated by ISC from a third triplet state between the GS and ES. ZPL dipole transitions between the GS and the \spinstate{3}{A}{2u} are forbidden, but can be driven into the vibronic sideband by emission of \spinstate{}{A}{2g} phonons. In both cases the given singlet characters are examples; however, there can be no dipole transition between singlets in order for the upper singlet to be an effective shelving state, as observed in temperature-dependent PL measurements \cite{DHaenens-Johansson2011}. In comparison to the well-understood model for \NVminus{} \cite{Doherty2011}, our model must also account for spin polarization at below-resonant energies.}
	\label{fig:spin_pol_mechanism}
\end{figure}

We now consider the source of the spin polarization. In \NVminus{}, the electronic structure relevant to spin polarization is described by the molecular orbitals (MO) \oneelec{a_1e}, with the \spinstate{3}{A}{2} GS generated by the configuration \oneelec{a_1^2 e^2} \cite{Doherty2011}. Spin polarization occurs by intersystem crossing (ISC) from the \spinstate{3}{E}{} ES (\oneelec{a_1^1 e^3}) into a pair of singlets (\spinstate{1}{A}{1}, \spinstate{1}{E}{}) arising from the same orbital configuration as the GS \cite{Goldman2015}. In \SiVneutral{}, the \spinstate{3}{A}{2g} GS is described by the MO configuration \oneelec{a_{1g}^2 a_{2u}^2 e_u^4 e_g^2}, which also produces two singlet states \spinstate{1}{A}{1g} \& \spinstate{1}{E}{g}. The configuration \oneelec{a_{1g}^2 a_{2u}^2 e_u^3 e_g^3} is responsible for the \spinstate{3}{A}{1u} ES and additional states \spinstate{1}{A}{1u}, \spinstate{1}{A}{2u}, \spinstate{1}{E}{u}, \spinstate{3}{A}{2u} and \spinstate{3}{E}{u}. The multitude of available states suggests the possibility of spin-orbit (SO) mediated ISC mechanisms, similar to \NVminus{} and other defects in diamond and SiC \cite{Ivady2015}: any model for \SiVneutral{} must account both for PL quenching at low temperature \cite{DHaenens-Johansson2011} and spin polarization generated by sub-ZPL excitation. Two of the possible energy level schemes which are consistent with experiment are given in Figure~\ref{fig:spin_pol_mechanism}, both based on ISC between singlet and triplet states. Sub-ZPL polarization is generated in (a) by pumping directly into the singlet state, which becomes weakly-allowed due to SO effects in the ground state. In (b), no-phonon dipole transitions from the \spinstate{3}{A}{2g} GS to the \spinstate{3}{A}{2u} ES are forbidden, but transitions into the vibronic sideband would be possible by emission of an \spinstate{}{A}{2g} phonon. Such an absorption would have no ZPL and the broad band may be difficult to detect. In this model, the spin polarization is generatred by ISC from the \spinstate{3}{A}{2u} triplet to a singlet (\spinstate{1}{E}{u}), and not from the ES involved in the \SI{946}{\nano\meter} ZPL. Resonant excitation to the \spinstate{3}{A}{1u} level would nevertheless result in spin polarization via non-radiative transitions from \spinstate{3}{A}{1u} to \spinstate{3}{A}{2u}. In both models, transitions between the singlet states must be dipole forbidden in order for the upper state to be an effective shelving state \cite{DHaenens-Johansson2011}. Detailed calculation of level energies and ordering is beyond the scope of the present work; nevertheless, the model emphasises that there are different possible polarization mechanisms. When pumping at the photoconductivity threshold or below (\SI{<830}{\nano\meter} \cite{Allers1995}), additional mechanisms are expected to occur: further work is required to understand the electronic structure and spin polarization mechanism of \SiVneutral{}.

Bulk spin polarization and long spin lifetimes at \SI{30}{\kelvin} and below, combined with a high Debye-Waller factor and infrared emission, establish \SiVneutral{} as a defect which demands further study. In particular, optical stress measurements would unambiguously identify the excited state of the \SI{946}{\nano\meter} ZPL and in turn aid in the interpretation of the observed echo-detected \SI{134}{\giga\hertz} ZPL splitting. Additionally, demonstration of spin-dependent photoluminescence contrast would enable the rapid determination of center properties and enable its exploitation as e.g.\ a remote temperature sensor. The ability to efficiently spin-polarize \SiVneutral{} at wavelengths which do not affect \NVminus{} opens the possibility of protocols which use \NVminus{} as a control/readout mechanism but where multiple qubits can be initialized independently of the \NVminus{} center and within the same optically-addressed volume. The \SI{946}{\nano\meter} wavelength falls within the \SI{980}{\nano\meter} band, where downconversion to the telecoms \SI{1550}{\nano\meter} wavelength has already been demonstrated \cite{Li2015a}: if future studies detect ODMR from \SiVneutral{} then it will be a compelling candidate for long-range quantum communication networks. 

We thank TOPTICA Photonics AG for the use of the CTL 950 tuneable laser. This work was supported by EPSRC grants EP/J500045/1 \& EP/M013243/1.

\pagebreak
\widetext
\begin{center}
\textbf{\large Supplemental Material}
\end{center}
\setcounter{equation}{0}
\setcounter{figure}{0}
\setcounter{table}{0}
\setcounter{page}{1}
\makeatletter
\renewcommand{\theequation}{S\arabic{equation}}
\renewcommand{\thefigure}{S\arabic{figure}}
\renewcommand{\thetable}{S\arabic{table}}
\renewcommand{\thepage}{S\arabic{page}}
\renewcommand{\bibnumfmt}[1]{[S#1]}
\renewcommand{\citenumfont}[1]{S#1}

\section{Calculation of bulk spin polarization percentage}
When considering the spin polarization of the $^3A_2$ triplet ground state it is necessary to examine the spin populations of the the $m_{S}=0$ and $\pm1$ spin levels. The fractional spin population of the $i^{\text{th}}$ level is given by
\begin{equation}
p_{i} = N_{i}/N
\end{equation}
where $N_{i}$ denotes the number of spins in the $i^{\text{th}}$ spin level and $N$ is the total number of spins in the triplet state. For a three level system,
\begin{equation} \label{eqn:poptotal}
p_{+1} + p_{0} + p_{-1} = 1
\end{equation}
where $p_{+1}$, $p_{0}$ and $p_{-1}$ are the fractional populations of the $m_{S}=+1$, 0 and $-1$ levels, respectively. The EPR signal intensity is proportional to the population difference between levels
\begin{eqnarray}
\eta_{+1,0} &= p_{0} - p_{+1}\\
\eta_{0,-1} &= p_{-1} - p_{0}\;.
\end{eqnarray}
In the absence of optical pumping the population of the $i^{\text{th}}$ spin level is given by Boltzmann statistics according to
\begin{equation}
p_{i}=\frac{e^{-E_{i}(B)/k_{B}T}}{\sum\limits_{i=1}^{N}e^{-E_{i}(B)/k_{B}T}}\;,
\end{equation}
where $E_{i}(B)$ is the energy of the $i^{\text{th}}$ spin level in a magnetic field $B$, $k_{B}$ is the Boltzmann constant, $T$ is the sample temperature and $N$ is the total number of spin states.

Optical pumping induces spin polarization and hence the occupation probabilities are no longer in thermal equilibrium (i.e.\ non-Boltzmann). It is possible to calculate the difference in the occupation probabilities under illumination by considering the experimentally determined intensities of the low and high field resonance lines:
\begin{equation} \label{eqn:etalight}
{\eta}^{l} = {\eta}^{d}\frac{I^{l}}{I^{d}}\;,
\end{equation}
where $d$ and $l$ refer to the sample in the dark and light, respectively. $I$ is positive if the transition is in absorption and negative if the transition is in emission. However, Eq.~\ref{eqn:etalight} only applies when the transitions are not microwave-power saturated: in our experiments it was not possible to measure in non-saturated conditions at \SI{10}{\kelvin} due to the long spin $T_1$. The unsaturated value for $I^{d}_{T}$ ($I^{d}$ at temperature $T$) was instead calculated using the following relationship, based on the experimental value for $I^{d}_{RT}$ ($I^{d}$ at room temperature) and the theoretical occupation probability difference between the spin levels at the different temperatures in the dark:
\begin{equation}
I^{d}_{T} = I^{d}_{RT} \frac{{\eta}^{d}_{T}}{{\eta}^{d}_{RT}}.
\end{equation}
Solving for the occupation probabilities of the $m_{S}=+1$, 0 and -1 spin levels under optical illumination one finds
\begin{eqnarray}
p_{+1}^{l}&=&\frac{1}{3} \left[1-2\left(\eta^{d}_{RT}\frac{I^{l}_{T}}{I^{d}_{RT}}\right)_{+1,0} - \left(\eta^{d}_{RT}\frac{I^{l}_{T}}{I^{d}_{RT}}\right)_{0,-1}\right] \\
p_{0}^{l}&=&\frac{1}{3} \left[1+\left(\eta^{d}_{RT} \frac{I^{l}_{T}}{I^{d}_{RT}}\right)_{+1,0} - \left(\eta^d_{RT}\frac{I^{l}_{T}}{I^{d}_{RT}} \right)_{0,-1} \right] \label{eqn:p0l}\\
p_{-1}^{l}&=&\frac{1}{3} \left[1 + \left(\eta^{d}_{RT} \frac {I^{l}_{T}} {I^{d}_{RT}} \right)_{+1,0} + 2\left(\eta^{d}_{RT}\frac{I^{l}_{T}}{I^{d}_{RT}}\right)_{0,-1} \right]
\end{eqnarray}

The degree of optical spin polarization $\xi$ is defined as
\begin{equation} \label{eqn:xi}
\xi = \frac{p^{l}_{0}-p^{d}_{0}}{p_{+1}^{d}+p_{-1}^{d}}
\end{equation}
Thus, $\xi=100\%$ occurs when all the spins are in the $m_{S}=0$ spin level and $p_{0}^{l}$\,$=$\,$p_{+1}^{d}$\,$+$\,$p_{0}^{d}$\,$+$\,$p_{-1}^{d}$, with all values calculated at the same temperature $T$. At the other extreme, when $p_{0}^{l}=p_{0}^{d}$ the system is not optically spin polarised and $\xi=0\%$.

\section{Broadening of EPR linewidth at high temperature}
\begin{figure}[bth]
	\centering
	\includegraphics[width=0.6\columnwidth]{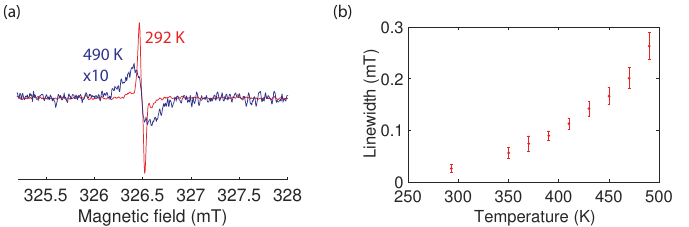}
	\caption{(a) Measured \SiVneutral{} spectra of the $m_{S}=0\leftrightarrow1$ transition at \num{292} and \SI{490}{\kelvin} with the magnetic field $B\|\hkl<110>$. The high temperature spectrum has been scaled for clarity. (b) Temperature-dependence of \SiVneutral{} linewidth above room temperature, showing significant broadening at elevated temperatures.}
	\label{fig:linewidth}
\end{figure}

At elevated temperatures (above approximately \SI{300}{\kelvin}), significant broadening of the EPR linewidth of \SiVneutral{} is observed [Fig.~\ref{fig:linewidth}]. The linewidth can be used as an indirect measurement of $T_1$ in the limit that $T_2$ is $2T_1$-limited \cite{S_Slichter1990}. The method is indirect as the linewidth also includes contributions from e.g.\ unresolved hyperfine couplings to distant \Cthir{}, which will broaden the line without affecting $T_1$. The linewidth-derived data included in Fig.~4(c) of the main text have been obtained as follows:
\begin{packed_itemize}
	\item Data were fit to extract raw linewidth in mT at each temperature
	\item Linewidth was converted to MHz using $\gamma_e=\SI{28.025}{\mega\hertz\per\milli\tesla}$
	\item At room temperature, $T_2=\SI{900}{\nano\second}$ is within a factor of $4$ of $2T_1=\SI{3.9}{\micro\second}$ ($\SI{256}{\kilo\hertz}$) and $T_1$ is changing rapidly. Therefore we assume that at temperatures of \SI{300}{\kelvin} and above, $T_2$ is limited by $2T_1$. 
	\item All linewidth data are offset by \SI{-250}{\kilo\hertz} so that the directly-measured $T_1$ and linewidth-derived values match at room temperature: this offset is reasonable, as typical linewidths due to unresolved \Cthir{} couplings are of the order of \SI{700}{\kilo\hertz} \cite{S_Clevenson2015}, and will increase the derived $1/T_1$ value.
\end{packed_itemize}

\section{Polarization comparisons}

\begin{figure}
	\centering
	\includegraphics[width=\columnwidth]{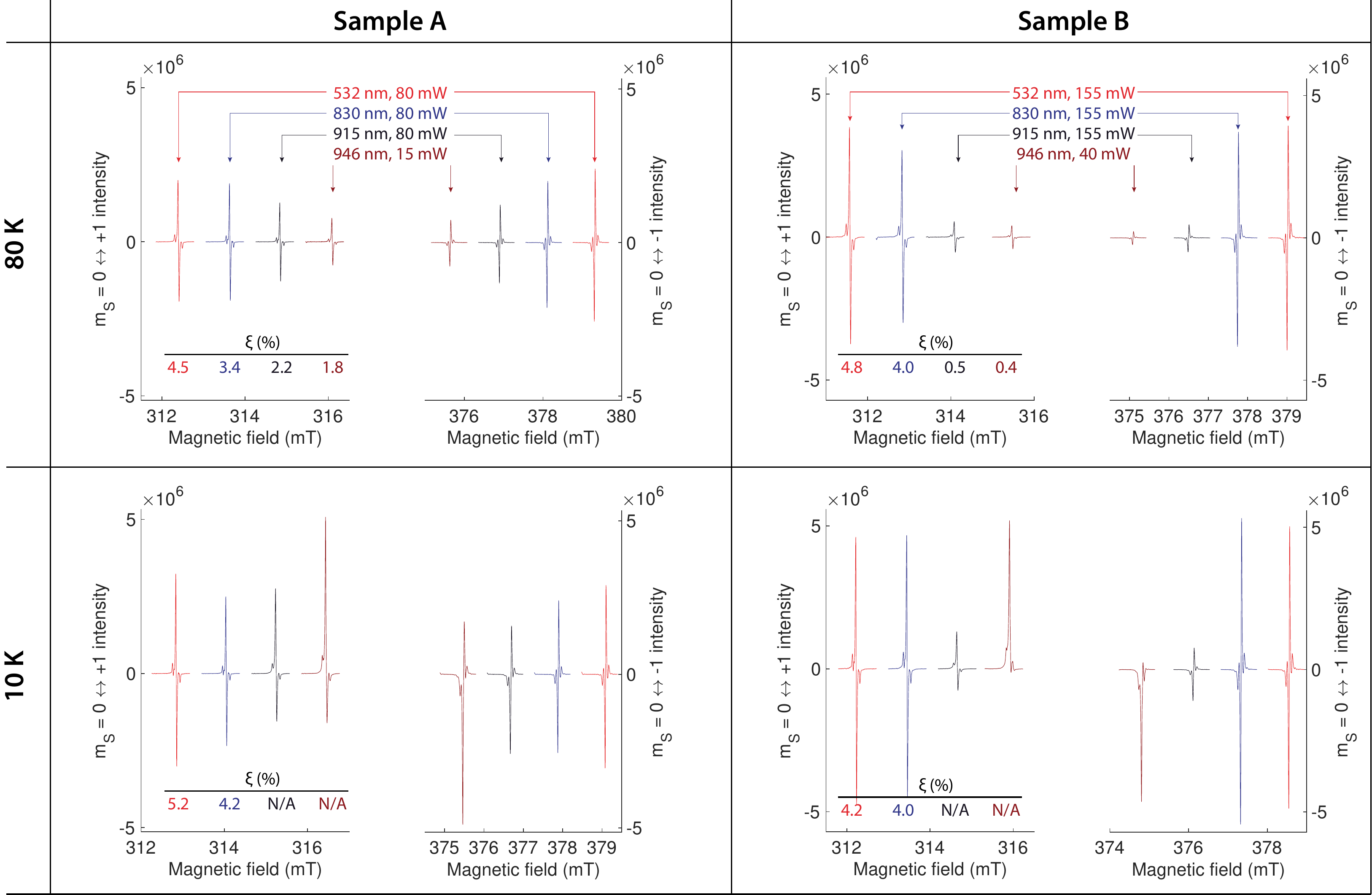}
	\caption{Raw CW EPR spectra for both samples at different temperatures and under optical excitation from different wavelengths. Available optical power was kept constant at both temperatures for each sample, but was different between samples due to experimental constraints. In all measurements the spin polarization was linear in optical power. Spectra have been offset in magnetic field for clarity. The asymmetric lineshapes visible at \SI{10}{\kelvin} under excitation at \num{915} and \SI{946}{\nano\meter} are due to microwave saturation of the transition: the same effect (a result of long $T_1$ at these temperatures in the absence of optical excitation) prohibited collection of dark reference spectra at low temperature. Bulk spin polarization values ($\xi$) have been given for spectra which are unsaturated --- they have not been normalized in any way.}
	\label{fig:EPR_spectral_comparison}
\end{figure}

\begin{figure}
	\centering
	\includegraphics[width=0.6\columnwidth]{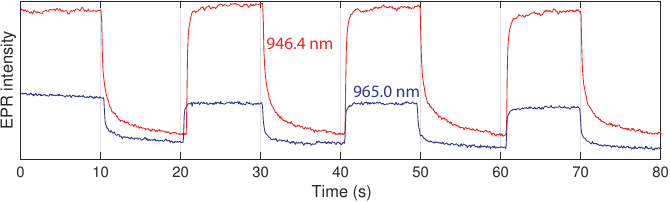}
	\caption{CW-detected EPR measurements of the high-field (emission) line of sample~A. The applied magnetic field is static, and the intensity of the $m_{S}=0\leftrightarrow-1$ transition is monitored as a function of time, with the laser switched manually on and off every \SI{10}{\second}. Two scans are shown, corresponding to ZPL-resonant (\SI{946.4}{\nano\meter}) and above-ZPL excitation.}
	\label{fig:on_off}
\end{figure}

Both samples A \& B were measured using CW EPR at four different wavelengths and two different temperatures. The results are given in \ref{fig:EPR_spectral_comparison}. It is clear that there are differences in the polarization efficiency at different wavelengths between the two samples, specifically the marked difference in behaviour at the higher energy (\num{830} and \SI{532}{\nano\meter}) and lower energy (\num{915} and \SI{946}{\nano\meter} excitation): as \SI{830}{\nano\meter} (\SI{1.4}{\electronvolt}) is the threshold for photocurrent from \SiVneutral{} \cite{S_Allers1995}, it seems likely that the two samples are undergoing different degrees of charge transfer. Nevertheless, the visibility of polarization at all wavelengths suggests either direct pumping into a higher-energy state which decays into the spin-polarization path, or that ionization of $\mathrm{SiV^{-}/SiV^{+}}$ can directly produce \SiVneutral{} in a spin-polarized state. The situation is complicated by below-ZPL spin polarization [Fig.~3, Fig.~\ref{fig:on_off}]: further investigation is required to understand the spin polarization and charge transfer behavior of this system.

\end{document}